\documentclass[aps,prl,reprint,amsmath,amssymb,longbibliography,nobibnotes]{revtex4-2}

  \usepackage[T1]{fontenc}
  \usepackage{lmodern}

\usepackage[utf8]{inputenc}
\usepackage{newunicodechar}
\newunicodechar{–}{--}
\usepackage{graphicx}
\usepackage{bm}
\usepackage[hidelinks]{hyperref}

\hypersetup{
  pdftitle={Extrinsic Bulk Photovoltaic Effect from Asymmetric Defect Scattering},
  pdfauthor={Zixun Wang and Andrew M. Rappe}
}

  \newcommand{\ManuscriptSection}[1]{\par\noindent\textit{#1.---}\ignorespaces}

  \newcommand{\SingleFigureWidth}{\linewidth}
  \newcommand{\WideFigureWidth}{0.83\linewidth}
  \newenvironment{widefigure}{\begin{figure*}[t]}{\end{figure*}}

\begin{document}

\title{Extrinsic Bulk Photovoltaic Effect from Asymmetric Defect Scattering}

\author{Zixun Wang}
\author{Andrew M. Rappe}
\email[Contact author: ]{rappe@sas.upenn.edu}
\affiliation{Department of Chemistry, University of Pennsylvania,
Philadelphia, Pennsylvania 19104--6323, USA}

\begin{abstract}
The extrinsic bulk photovoltaic effect from asymmetric electron-defect scattering can make an important contribution to the total photocurrent and is highly diagnostic of defect type. We present an ab initio Boltzmann framework combining optical excitation with electron-phonon and symmetric and skew electron-defect scattering, and we apply it to monolayer GeS. Because the defects that generate the current also relax the anisotropy that sustains it, the response peaks at a finite defect fraction. Within the scattering-limited model, cooling to $0$~K shifts the peak position toward lower defect fractions and yields a maximum enhancement of more than six orders of magnitude at fixed defect fraction. At selected low-temperature defect fractions, the total extrinsic current can exceed the intrinsic shift current by more than an order of magnitude, so that the defect fraction itself becomes a tunable parameter.
\end{abstract}

\maketitle

\ManuscriptSection{Introduction}
In noncentrosymmetric, single-phase materials, incident light can generate a direct photocurrent in the bulk without a junction~\cite{vonBaltzKraut1981,SturmanFridkin1992,BelinicherSturman1980,TanReview2016,YoungZhengRappe2012}. This bulk photovoltaic effect (BPVE) is attributed to several primary mechanisms~\cite{DaiRappeReview2023,Sturman2020,Burger2019,Burger2020}. First-principles calculations have largely focused on intrinsic mechanisms associated with photoexcitation~\cite{WangRappe2015,SchanklerPiezophotovoltaic2021,TanRappeTopology2016}, including the shift current and injection current~\cite{SipeShkrebtii2000,YoungRappe2012,FeiTanRappe2020}, as well as the asymmetric photoexcitation processes including the ballistic current~\cite{Dai2021,DaiRappeElectronHole2021}. More recently, intrinsic contributions from carrier relaxation and recombination~\cite{BelinicherIvchenkoSturman1982,ZhuAlexandradinata2024} have also been modeled from first principles~\cite{Schankler2026,YuPing2026}. However, extrinsic mechanisms such as asymmetric defect scattering, while identified in early kinetic theories of the BPVE~\cite{BelinicherSturman1980} and indicated by experiments to strongly reshape photoexcited carrier distributions~\cite{Ganguly2025}, have not, to our knowledge, been evaluated quantitatively from first principles in real materials. Consequently, the magnitude and controllability of extrinsic photocurrents remain largely unknown.

In this work, we develop a first-principles Boltzmann transport equation (BTE) framework to compute the leading extrinsic BPVE generated by electron-defect skew scattering. By combining ab initio optical excitation, electron-phonon scattering, and both symmetric and asymmetric electron-defect scattering within an iterative BTE solver, we evaluate the steady photocurrent beyond the relaxation-time approximation. Furthermore, we derive a minimal analytical model that reveals the competition between skew conversion and defect-induced relaxation, predicting a crossover from linear to inverse scaling of photocurrent with defect fraction and identifying a finite optimal defect fraction. For monolayer GeS containing either of two representative point defects, we find that this optimal defect fraction is strongly temperature-dependent and that the extrinsic current can significantly exceed the intrinsic shift current in the cold, dilute regime.

\ManuscriptSection{Theory}
In a crystal lacking inversion symmetry, electron-defect scattering can contain an antisymmetric component. Denoting by
$W_{n\mathbf{k},n'\mathbf{k}'}$
the transition rate from the Bloch state
$|n\mathbf{k}\rangle$
to
$|n'\mathbf{k}'\rangle$,
in general,
\begin{equation}
W_{n\mathbf{k},n'\mathbf{k}'}
\neq
W_{n'\mathbf{k}',n\mathbf{k}}.
\label{eq:asymmetry}
\end{equation}
Inversion asymmetry permits this contribution, subject to the remaining crystal symmetries~\cite{BelinicherSturman1980,Deyo2009,Sukhachov2026}.

Within the leading skew-scattering approximation~\cite{Deyo2009,IsobeXuFu2020,Sukhachov2026}, we expand the single-defect transition matrix through second order in the defect perturbation potential $V_d$,
\begin{equation}
\mathcal{T}_{n\mathbf{k},n'\mathbf{k}'}
\simeq
U_{n\mathbf{k},n'\mathbf{k}'}
+
\sum_{n''\mathbf{k}''}
U_{n\mathbf{k},n''\mathbf{k}''}
G^{R}_{n''\mathbf{k}''}(\epsilon_{n\mathbf{k}})
U_{n''\mathbf{k}'',n'\mathbf{k}'},
\label{eq:Tmatrix}
\end{equation}
where
\begin{equation}
U_{n\mathbf{k},n'\mathbf{k}'}
=
\langle n\mathbf{k}|V_d|n'\mathbf{k}'\rangle
\label{eq:U}
\end{equation}
is the electron-defect matrix element, $\epsilon_{n\mathbf{k}}$ is the Bloch-state energy, and $G^{R}_{n''\mathbf{k}''}(\epsilon)$ is the retarded electronic Green's function.

Retaining only the first-order term in Eq.~\eqref{eq:Tmatrix} yields the symmetric defect-scattering kernel
\begin{equation}
W^{\mathrm{sym}}_{n\mathbf{k},n'\mathbf{k}'}
=
\frac{2\pi}{\hbar}
\left|
U_{n\mathbf{k},n'\mathbf{k}'}
\right|^2
\delta
\left(
\epsilon_{n\mathbf{k}}-\epsilon_{n'\mathbf{k}'}
\right).
\label{eq:symmetric_rate}
\end{equation}
We define the asymmetric, or skew, component as
\begin{equation}
W^{\mathrm{skew}}_{n\mathbf{k},n'\mathbf{k}'}
=
\frac{1}{2}
\left(
W_{n\mathbf{k},n'\mathbf{k}'}
-
W_{n'\mathbf{k}',n\mathbf{k}}
\right).
\label{eq:skew_definition}
\end{equation}
Upon substitution of the $T$-matrix expansion, the principal-value contribution cancels by Hermiticity of $V_d$, leaving the on-shell contribution
\begin{align}
W^{\mathrm{skew}}_{n\mathbf{k},n'\mathbf{k}'}
={}&
\frac{(2\pi)^2}{\hbar}
\sum_{n''\mathbf{k}''}
\mathrm{Im}
\left[
U_{n\mathbf{k},n''\mathbf{k}''}
U_{n''\mathbf{k}'',n'\mathbf{k}'}
U_{n'\mathbf{k}',n\mathbf{k}}
\right]
\nonumber\\
&\times
\delta
\left(
\epsilon_{n\mathbf{k}}-\epsilon_{n'\mathbf{k}'}
\right)
\delta
\left(
\epsilon_{n\mathbf{k}}-\epsilon_{n''\mathbf{k}''}
\right).
\label{eq:skew_rate}
\end{align}
This rate satisfies
\begin{equation}
W^{\mathrm{skew}}_{n\mathbf{k},n'\mathbf{k}'}
=
-
W^{\mathrm{skew}}_{n'\mathbf{k}',n\mathbf{k}}.
\label{eq:skew_antisymmetry}
\end{equation}
Here, ``skew scattering'' denotes the leading antisymmetric rate, Eq.~\eqref{eq:skew_rate}, cubic in the defect potential~\cite{Deyo2009,IsobeXuFu2020,Sukhachov2026}. For dilute, uncorrelated defects, the single-defect kernels in Eqs.~\eqref{eq:symmetric_rate} and \eqref{eq:skew_rate} yield rates linear in the defect fraction $n_{\mathrm{imp}}$~\cite{Lu2019,Sukhachov2026}.

Together with intrinsic electron-phonon scattering, these processes define the linear collision operator~\cite{Bernardi2016,Deyo2009}
\begin{equation}
L
=
L^{\mathrm{ph}}
+
L^{\mathrm{sym}}
+
L^{\mathrm{skew}},
\label{eq:L}
\end{equation}
where $L^{\mathrm{ph}}$, $L^{\mathrm{sym}}$, and $L^{\mathrm{skew}}$ denote the inelastic electron-phonon, elastic symmetric defect, and elastic skew defect collision operators, respectively~\cite{Giustino2017,Lu2019,Deyo2009}.

Under continuous-wave illumination in the low-injection limit, we treat the conduction-band electron and valence-band hole sectors independently. We neglect Pauli blocking and retain the collision term to first order in these dilute carrier occupations. For either carrier sector, the photoinduced distribution obeys~\cite{Bernardi2016,Zhou2021}
\begin{equation}
\partial_t\delta f_{n\mathbf{k}}
=
G_{n\mathbf{k}}
-
(L\delta f)_{n\mathbf{k}},
\label{eq:BTE}
\end{equation}
where $\delta f_{n\mathbf{k}}\ge0$ denotes the corresponding photoexcited
electron or hole distribution, $G_{n\mathbf{k}}$ is the ab initio optical
generation rate, and
$(L\delta f)_{n\mathbf{k}}
=\sum_{n'\mathbf{k}'}
\big[W_{n\mathbf{k},n'\mathbf{k}'}\,\delta f_{n\mathbf{k}}
-W_{n'\mathbf{k}',n\mathbf{k}}\,\delta f_{n'\mathbf{k}'}\big]$
is the out-minus-in collision term built from the electron-phonon rates and Eqs.~\eqref{eq:symmetric_rate} and \eqref{eq:skew_rate}~\cite{Jacoboni2010,Bernardi2016}.

Because carrier recombination is omitted, scattering alone cannot remove the photoexcited carriers~\cite{Jacoboni2010,Bernardi2016}. Summing Eq.~\eqref{eq:BTE} over the states within either carrier sector gives
\begin{equation}
\frac{d\mathcal{N}}{dt}
=
R_{\mathrm{gen}},
\label{eq:number_growth}
\end{equation}
where
\begin{equation}
\mathcal{N}(t)
=
\sum_{n\mathbf{k}}
\delta f_{n\mathbf{k}}(t),
\qquad
R_{\mathrm{gen}}
=
\sum_{n\mathbf{k}}
G_{n\mathbf{k}}
\label{eq:number_definition}
\end{equation}
are the photoexcited electron or hole population and its corresponding generation rate. The electron and hole populations grow together under interband excitation, while the total number of physical electrons remains conserved.  As shown next, the part of the distribution produced
by skew scattering, which carries the current, can reach a steady state even while the total
carrier population in both sectors is still growing.

To separate the growing population from the current-carrying component~\cite{SupplementalMaterial}, we write
\begin{equation}
\delta f(t)
=
a(t)\phi_0
+
\widetilde{f}(t),
\label{eq:decomposition}
\end{equation}
where $\phi_0$ is the population-conserving zero mode satisfying
$L\phi_0=0$,
$a(t)$ is its time-dependent amplitude, and
$\widetilde{f}$ has zero net carrier number. The secular population growth is confined to $a(t)$, while
$\widetilde{f}$ relaxes to a finite steady state
$\widetilde{f}^{\mathrm{ss}}$.

If $\phi_0$ is even under time reversal, the odd band velocity
$\mathbf{v}_{n\mathbf{k}}$ ensures that
\begin{equation}
\sum_{n\mathbf{k}}
\mathbf{v}_{n\mathbf{k}}
\phi_{0,n\mathbf{k}}
=
0.
\label{eq:zero_current}
\end{equation}
Under these conditions, the growing population carries no current and $\widetilde{f}^{\mathrm{ss}}$ determines the steady photocurrent~\cite{SupplementalMaterial,Norris1997}.

Two experimentally accessible parameters govern the skew photocurrent: the defect fraction per sulfur site $n_{\mathrm{imp}}$ and the temperature $T$. To describe their effects, we reduce the BTE to an angular-harmonic model~\cite{AshcroftMermin1976,Koniakhin2014,SupplementalMaterial}. On each energy shell, we expand the distribution in the polar angle $\theta$ of $\mathbf{k}$:
\begin{equation}
\delta f(\theta)
=
\sum_m f_m e^{im\theta},
\label{eq:harmonics}
\end{equation}
where $m$ is the angular-harmonic index and $f_m$ is the corresponding harmonic amplitude.

Within this minimal isotropic model, linearly polarized excitation generates an isotropic $m=0$ population and an $m=2$ optical alignment. Neither carries current directly, as the photocurrent resides in the odd $m=1$ sector~\cite{SupplementalMaterial}. Moreover,
\begin{equation}
L^{\mathrm{skew}}f_0=0,
\label{eq:isotropic_skew}
\end{equation}
meaning that the isotropic population cannot generate an $m=1$ component~\cite{SupplementalMaterial}. The skew photocurrent is therefore produced by conversion of the current-free $m=2$ alignment into the current-carrying $m=1$ component.

Let $\tau_m$ denote the relaxation time of angular harmonic $m$~\cite{AshcroftMermin1976,Koniakhin2014}, where
\begin{equation}
\tau_m^{-1}
=
\Gamma_m(T)
+
\alpha_mn_{\mathrm{imp}},
\label{eq:tau}
\end{equation}
with $\Gamma_m(T)$ being the phonon-induced relaxation rate and $\alpha_m$ representing the elastic defect-relaxation rate per unit defect fraction~\cite{SupplementalMaterial}. As mentioned above, the skew-scattering rate is likewise linear in $n_{\mathrm{imp}}$. Denoting the $m=2$ component of the optical generation rate by $G_2$, the optical alignment becomes
\begin{equation}
f_2
=
\frac{G_2}
{\Gamma_2(T)+\alpha_2n_{\mathrm{imp}}}.
\label{eq:f2}
\end{equation}

Acting on the $m=2$ alignment, skew scattering generates a source $S_2n_{\mathrm{imp}}f_2$ for the current-carrying $m=1$ harmonic, where $S_2$ is the skew-conversion coefficient. Balancing this source against relaxation gives the photocurrent density~\cite{SupplementalMaterial}
\begin{equation}
J(n_{\mathrm{imp}},T)
\propto
\frac{
S_2G_2n_{\mathrm{imp}}
}{
\left[
\Gamma_1(T)+\alpha_1n_{\mathrm{imp}}
\right]
\left[
\Gamma_2(T)+\alpha_2n_{\mathrm{imp}}
\right]
}.
\label{eq:minimal_model}
\end{equation}
At fixed illumination intensity, the skew conductivity shares this dependence on $n_{\mathrm{imp}}$ and $T$. Therefore, increasing the defect fraction strengthens the skew conversion but simultaneously diminishes the current-carrying and alignment lifetimes.

In the phonon-dominated dilute regime,
$\alpha_mn_{\mathrm{imp}}\ll\Gamma_m(T)$,
Eq.~\eqref{eq:minimal_model} reduces to
$J\propto n_{\mathrm{imp}}$. Conversely, once both relevant harmonics become defect-dominated,
$\alpha_mn_{\mathrm{imp}}\gg\Gamma_m(T)$,
the current obeys
$J\propto n_{\mathrm{imp}}^{-1}$, making the response self-limiting. If the crossover fractions
$\Gamma_1(T)/\alpha_1$ and
$\Gamma_2(T)/\alpha_2$
are well separated, an intermediate shoulder or plateau emerges when only one harmonic is defect-limited.

Maximizing the magnitude of Eq.~\eqref{eq:minimal_model} yields a unique optimal defect fraction for the modeled carrier channel,
\begin{equation}
n_{\mathrm{opt}}
=
\sqrt{
\frac{
\Gamma_1(T)\Gamma_2(T)
}{
\alpha_1\alpha_2
}
}
=
\sqrt{
\left(
\frac{\Gamma_1(T)}{\alpha_1}
\right)
\left(
\frac{\Gamma_2(T)}{\alpha_2}
\right)
}.
\label{eq:Nopt}
\end{equation}
Thus, $n_{\mathrm{opt}}$ is the geometric mean of the current-carrying and alignment crossover fractions. Comparable scales produce a sharp turnover; well-separated scales broaden the maximum.

Cooling reduces the phonon relaxation rates $\Gamma_m(T)$, shifting the crossover fractions and $n_{\mathrm{opt}}$ to lower defect fractions. The cooling enhancement therefore depends strongly on defect fraction: it persists while phonons control the relevant lifetimes but is suppressed once defect scattering dominates. Spontaneous phonon emission can still contribute to $\Gamma_m$ at $T=0$~K, so that electron-phonon scattering need not vanish~\cite{Giustino2017,Zhou2021}.

The first-principles BTE retains the full angular structure: anisotropic energy contours and momentum-dependent optical and scattering matrix elements in GeS can mix harmonics beyond the $m=0$ and $m=2$ sources, allowing extrema or sign reversals absent in Eq.~\eqref{eq:minimal_model}~\cite{SupplementalMaterial}.

\ManuscriptSection{Computational details}
We perform density functional theory (DFT) calculations using \textsc{Quantum ESPRESSO}~\cite{Giannozzi2009} with the Perdew-Burke-Ernzerhof (PBE) functional~\cite{Perdew1996} and norm-conserving pseudopotentials. Monolayer GeS has in-plane lattice constants
$a=3.666$~\AA{} and
$b=4.455$~\AA{} and a
$15$~\AA{} out-of-plane cell parameter. We use a
$50$~Ry wave-function cutoff and an
$8\times8\times1$ Monkhorst-Pack electronic wave-vector mesh~\cite{MonkhorstPack1976}. The resulting PBE band structure features an indirect gap of
$1.78$~eV and a minimum direct gap of
$1.91$~eV at the Brillouin-zone center $\Gamma$.

We compute phonons and electron-phonon matrix elements using density functional perturbation theory (DFPT)~\cite{Baroni2001} on an
$8\times8\times1$ phonon wave-vector $\mathbf{q}$ mesh. These are interpolated using maximally localized Wannier functions~\cite{MarzariVanderbilt1997,Giustino2007}
constructed from the four highest valence and three lowest conduction
bands, which contain all vertical transitions within the photon-energy
range considered here.

The electron-defect matrix elements
$U_{n\mathbf{k},n'\mathbf{k}'}$
are calculated with \textsc{Perturbo}~\cite{Zhou2021}. The defect perturbation $V_d$ is obtained from the difference between the Kohn-Sham potentials of defect-containing and pristine
$4\times4\times1$ supercells~\cite{Lu2019,Lu2020}. We study the sulfur vacancy $\mathrm{V_S}$ and isovalent selenium substitution $\mathrm{Se_S}$, with substantially different leading-order scattering strengths. For dilute, uncorrelated defects, we obtain the $n_{\mathrm{imp}}$ dependence by rescaling the single-defect symmetric and skew rates.

We compute $G_{n\mathbf{k}}$ from interband momentum matrix elements $\mathbf{p}_{nn'}(\mathbf{k})$ using modified \textsc{Wannier90}~\cite{Pizzi2020} and Fermi's golden rule for continuous-wave light polarized along $\hat{\mathbf e}=\hat{\mathbf y}$ at $\hbar\omega=2.0$--$4.0$~eV.

We solve Eq.~\eqref{eq:BTE} on a
$100\times100\times1$
electronic wave-vector mesh via fixed-point iteration accelerated with Anderson mixing~\cite{Anderson1965}. This retains momentum-dependent in-scattering in $L=L^{\mathrm{ph}}+L^{\mathrm{sym}}+L^{\mathrm{skew}}$ beyond the relaxation-time approximation. We solve electrons and holes independently, separating the particle-number zero mode from the steady-state current-carrying distribution.

We incorporate temperature explicitly through the phonon occupations in the microscopic electron-phonon scattering rates, reporting results at different temperatures. The $T=0$~K calculation retains spontaneous phonon-emission processes allowed for the photoexcited carriers. For direct comparison, the intrinsic shift-current conductivity of monolayer GeS is computed under identical conditions~\cite{YoungZhengRappe2012,Rangel2017}.

\ManuscriptSection{Results}
For $mm2$ symmetry with the polar axis along $y$~\cite{Rangel2017}, the mirror symmetry
$m_x:x\rightarrow -x$
forbids an in-plane current along $x$ under $y$-polarized illumination, while a current along $y$ is allowed. We therefore consider the
$\sigma^{yyy}$ response and define
\begin{equation}
K_y
=
\sigma^{yyy}E_y^2,
\label{eq:sigma_definition}
\end{equation}
where $K_y$ is the photocurrent per unit width and $E_y$ is the optical electric field amplitude. We denote
$\sigma^{yyy}$ by $\sigma_y$ and report it in
nm$\cdot\mu$A/V$^2$. This component aligns with the largest intrinsic shift-current response considered for monolayer GeS~\cite{Rangel2017}.

We apply the framework across defect fractions
$n_{\mathrm{imp}}=10^{-7}$--$10^{-2}$. Because the analytical model treats one carrier sector at a time, we first compute the electron [conduction-band (CB)] and hole [valence-band (VB)] contributions,
$\sigma_y^{\mathrm{CB}}$ and
$\sigma_y^{\mathrm{VB}}$,
independently. Unless noted, results are presented at
$\hbar\omega=3.5$~eV.

\begin{figure}[tbp]
\centering
\includegraphics[width=\SingleFigureWidth]{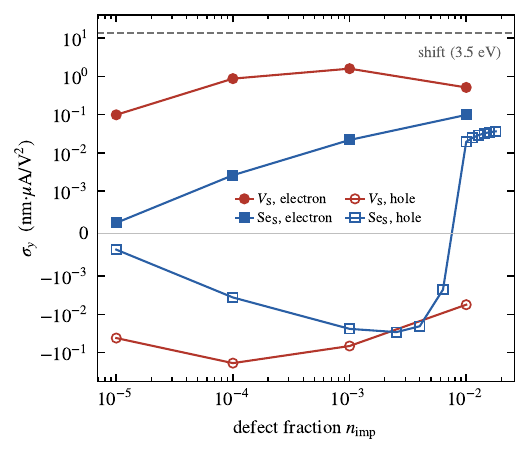}
\caption{
Extrinsic skew conductivity $\sigma_y$ of monolayer GeS at
$T=300$~K and $\hbar\omega=3.5$~eV as a function of defect fraction
$n_{\mathrm{imp}}$ per sulfur site, for
$\mathrm{V_S}$ (red circles) and
$\mathrm{Se_S}$ (blue squares).
Filled and open symbols denote the electron and hole channels, respectively.
The dashed line marks the intrinsic shift-current conductivity at the same photon energy.
The symmetric-logarithmic ordinate displays the zero crossing of the
$\mathrm{Se_S}$ hole response near
$n_{\mathrm{imp}}\simeq7\times10^{-3}$.
}
\label{fig:fraction}
\end{figure}

Figure~\ref{fig:fraction} illustrates the defect-fraction dependence of the steady-state skew current response at
$T=300$~K. For $\mathrm{V_S}$, both carrier channels exhibit the single-turnover structure predicted by Eq.~\eqref{eq:minimal_model}: the response initially scales linearly with
$n_{\mathrm{imp}}$
and is subsequently suppressed as defect scattering determines the relevant lifetimes. We identify peak positions near
$n_{\mathrm{opt}}\sim10^{-3}$
for electrons and
$n_{\mathrm{opt}}\sim10^{-4}$
for holes.

For $\mathrm{Se_S}$, the weaker defect scattering shifts the crossover toward larger
$n_{\mathrm{imp}}$,
resulting in a response that remains largely linear within the computed dilute range. The
$\mathrm{Se_S}$
hole channel exhibits a sign reversal, indicating additional microscopic skew-conversion pathways captured by the full BTE that are absent from the single-harmonic analytical model.

\begin{figure}[tbp]
\centering
\includegraphics[width=\SingleFigureWidth]{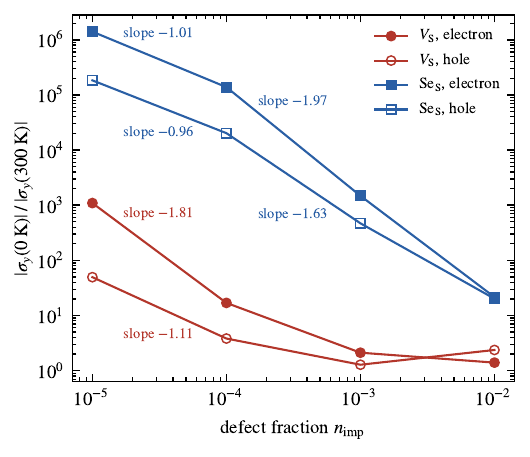}
\caption{
Cooling enhancement factor
$|\sigma_y(0\,\mathrm{K})|/
|\sigma_y(300\,\mathrm{K})|$
at $\hbar\omega=3.5$~eV as a function of defect fraction
$n_{\mathrm{imp}}$.
Symbols are as in Fig.~\ref{fig:fraction}.
This enhancement ranges from order unity in the defect-dominated part of the dilute regime to over
$10^{6}$ at low defect fractions.
Labels indicate logarithmic slopes between adjacent calculated points.
}
\label{fig:cooling}
\end{figure}

The cooling enhancement depends strongly on defect fraction [Fig.~\ref{fig:cooling}]. The enhancement factor
\[
\frac{|\sigma_y(0\,\mathrm{K})|}
{|\sigma_y(300\,\mathrm{K})|}
\]
ranges from order unity at the defect-dominated edge to over
$10^{6}$ at low defect fractions. In the phonon-dominated regime, where
$\tau_m\simeq\Gamma_m(T)^{-1}$,
the suppression of thermal phonons strongly amplifies the response.

Each factor $\Gamma_m(T)+\alpha_m n_{\mathrm{imp}}$ in
Eq.~\eqref{eq:minimal_model} turns from constant to linear at
$n_{\mathrm{imp}}=\Gamma_m(T)/\alpha_m$, and the $0$~K thresholds lie below
the $300$~K ones, so the logarithmic slope of the enhancement steepens from
$0$ to $-1$ to $-2$ as the two $0$~K thresholds are crossed and relaxes
back as the $300$~K thresholds follow. The segment slopes in
Fig.~\ref{fig:cooling} reflect this trend: $\mathrm{Se_S}$ steepens from
$-1.01$ to $-1.97$ (electrons) and $-0.96$ to $-1.63$ (holes) between
$10^{-5}$ and $10^{-3}$, whereas $\mathrm{V_S}$, whose thresholds are roughly an order of
magnitude lower, already shows $-1.81$ (electrons) and $-1.11$ (holes) in the
first decade and flattens above $10^{-4}$.

The measurable skew conductivity is the signed sum
\begin{equation}
\sigma_y^{\mathrm{tot}}
=
\sigma_y^{\mathrm{CB}}
+
\sigma_y^{\mathrm{VB}}.
\label{eq:total_sigma}
\end{equation}
As shown in Fig.~\ref{fig:lowT}(a), cooling shifts the peak position toward lower defect fractions. For
$\mathrm{V_S}$,
the electron-channel peak shifts from
$n_{\mathrm{imp}}\sim10^{-3}$
at
$300$~K
to
$n_{\mathrm{imp}}\sim10^{-6}$
at
$0$~K,
consistent with the trend predicted by Eq.~\eqref{eq:Nopt}.

\begin{widefigure}
\centering
\includegraphics[width=\WideFigureWidth]{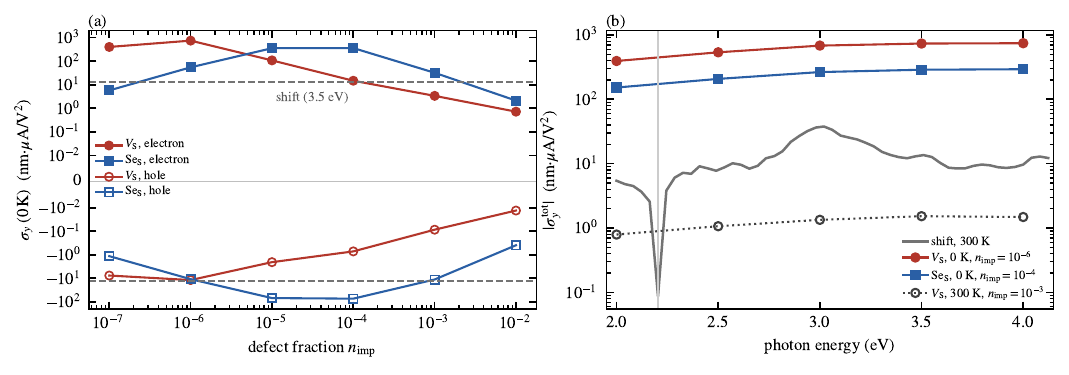}
\caption{
Low-temperature response of monolayer GeS.
(a) Skew conductivity $\sigma_y$ at
$T=0$~K and $\hbar\omega=3.5$~eV as a function of defect fraction
$n_{\mathrm{imp}}$, for
$\mathrm{V_S}$ and
$\mathrm{Se_S}$.
Filled and open symbols denote the electron and hole channels, respectively.
Dashed lines mark $\pm$ the intrinsic shift-current conductivity at the same photon energy.
(b) Total extrinsic response
$|\sigma_y^{\mathrm{tot}}|
=
|\sigma_y^{\mathrm{CB}}+\sigma_y^{\mathrm{VB}}|$
at selected defect fractions held fixed across five photon energies:
the $T=0$~K response of
$\mathrm{V_S}$ at
$n_{\mathrm{imp}}=10^{-6}$
and of
$\mathrm{Se_S}$ at
$n_{\mathrm{imp}}=10^{-4}$,
the $T=300$~K
$\mathrm{V_S}$ response at
$n_{\mathrm{imp}}=10^{-3}$ (gray, dotted),
and the intrinsic shift-current spectrum (solid gray).
The selected fractions lie near the electron-channel peaks in the $3.5$~eV defect-fraction scans and are not reoptimized at each photon energy. The dip near $2.2$~eV marks the sign change of the shift-current response.
}
\label{fig:lowT}
\end{widefigure}

The absolute magnitude of this effect is substantial [Fig.~\ref{fig:lowT}(b)]. At
$300$~K at $n_{\mathrm{imp}}=10^{-3}$, the total extrinsic response provides a non-negligible contribution, reaching up to approximately
$15\%$
of the intrinsic shift-current magnitude across the calculated spectrum. At
$0$~K, the extrinsic contribution can substantially exceed the intrinsic shift current. At the selected defect fractions in Fig.~\ref{fig:lowT}(b), the
$\mathrm{Se_S}$ and
$\mathrm{V_S}$ responses exceed the shift-current magnitude by factors of
$7$--$30$
and
$18$--$77$,
respectively. These large total responses occur at finite defect fractions near the electron-channel peaks, rather than in the perfectly pristine limit.

At the lowest temperatures and most dilute defect fractions, carrier relaxation becomes slow enough that recombination may provide an additional limiting time scale. In this regime, the present results should be interpreted as the scattering-limited response of the asymmetric-defect mechanism~\cite{SupplementalMaterial}. Adding carrier-recombination terms and occupation-dependent generation and scattering to the BTE would account for carrier loss and saturation, which can limit the enhancement and shift the optimal defect fraction.

\ManuscriptSection{Conclusion}
We extend first-principles BPVE calculations to photocurrents generated by asymmetric defect scattering, treating optical generation, phonon relaxation, and symmetric and skew defect scattering within a common Boltzmann framework. In monolayer GeS, competition between skew conversion and defect-induced relaxation produces a finite optimal defect fraction that shifts to lower values upon cooling. The low-temperature extrinsic response can exceed the intrinsic shift current. At fixed photon energy and illumination intensity, the minimal model predicts a single-channel photocurrent that rises linearly with defect fraction and decays inversely with it beyond the maximum. Measurements across defect fraction and temperature could test this competition. The contrasting dependence on defect fraction of the two defect types, including the sign reversal of the Se$_{\mathrm S}$ hole contribution at $300$~K, also suggests that photocurrent measurements could help distinguish defect types, provided that electron, hole, and intrinsic contributions can be resolved.

\begin{samepage}

\par\noindent\textit{Acknowledgments.---}\ignorespaces

This work was supported by the U.S. Department of Energy, Office of Science, Basic Energy Sciences, under Award No. DE-SC0024942. Computational support was provided by the National Energy Research Scientific Computing Center (NERSC), a U.S. Department of Energy, Office of Science User Facility located at Lawrence Berkeley National Laboratory, operated under Contract No. DE-AC02-05CH11231.
\par
\end{samepage}

\bibliographystyle{apsrev4-2}

\makeatletter
\immediate\write\@auxout{\string\citation{APSLayoutControl}}
\makeatother
\bibliography{references}

  \clearpage
  \onecolumngrid
  \let\SMBeginDocument\relax
  \let\SMEndDocument\relax

\newcommand{\bk}{\mathbf{k}}
\newcommand{\bv}{\mathbf{v}}
\newcommand{\bp}{\mathbf{p}}
\newcommand{\be}{\hat{\mathbf{e}}}
\newcommand{\Nimp}{n_{\mathrm{imp}}}
\newcommand{\Nopt}{n_{\mathrm{opt}}}
\newcommand{\Lsk}{L^{\mathrm{skew}}}
\newcommand{\Lsym}{L^{(S)}}
\newcommand{\Wsk}{W^{\mathrm{skew}}}
\renewcommand{\Im}{\operatorname{Im}}
\renewcommand{\Re}{\operatorname{Re}}

\SMBeginDocument

  \makeatletter
\let\SMcite\cite
\renewcommand{\cite}[1]{\begingroup\@fileswfalse\SMcite{#1}\endgroup}
\makeatother
\renewcommand{\citenumfont}[1]{S#1}
  \renewcommand{\bibnumfmt}[1]{[S#1]}

\begingroup
\normalfont\normalsize
\setcounter{section}{0}
\setcounter{subsection}{0}
\setcounter{equation}{0}
\setcounter{figure}{0}
\setcounter{table}{0}
\setcounter{secnumdepth}{2}
\renewcommand{\thesection}{S\arabic{section}}
\renewcommand{\thesubsection}{\thesection.\arabic{subsection}}
\renewcommand{\theequation}{S\arabic{equation}}
\renewcommand{\thefigure}{S\arabic{figure}}
\renewcommand{\thetable}{S\arabic{table}}
\renewcommand{\theHsection}{supplement.\arabic{section}}
\renewcommand{\theHsubsection}{supplement.\arabic{section}.\arabic{subsection}}
\renewcommand{\theHequation}{supplement.\arabic{equation}}
\renewcommand{\theHfigure}{supplement.\arabic{figure}}
\renewcommand{\theHtable}{supplement.\arabic{table}}
\makeatletter
\def\p@subsection{}
\def\p@subsubsection{}
\gdef\ps@supplement{  \def\@oddhead{\normalfont\normalsize\hfil\thepage}  \let\@evenhead\@oddhead
  \let\@oddfoot\@empty
  \let\@evenfoot\@empty
}
\renewcommand\section{\@startsection{section}{1}{\z@}  {2.8ex plus .8ex minus .3ex}{1.3ex plus .3ex}  {\normalfont\small\bfseries\centering}}
\renewcommand\subsection{\@startsection{subsection}{2}{\z@}  {2.3ex plus .6ex minus .2ex}{1.0ex plus .2ex}  {\normalfont\small\bfseries\centering}}
\makeatother
\pagestyle{supplement}
\thispagestyle{supplement}
\setlength{\emergencystretch}{1em}

\begin{center}
{\large Supplemental Material for\par}
\vspace{3pt}
{\normalsize\bfseries Extrinsic Bulk Photovoltaic Effect from Asymmetric Defect Scattering\par}
\vspace{8pt}
Zixun Wang and Andrew M. Rappe\par
\vspace{3pt}
{\small\itshape Department of Chemistry, University of Pennsylvania,\par
Philadelphia, Pennsylvania 19104--6323, USA\par}
\end{center}
\vspace{8pt}
\noindent
Section~\ref{sm:sec:steady} gives sufficient conditions for a finite steady
photocurrent in the Boltzmann equation without recombination. This result is used in the main text after
Eq.~(13). Section~\ref{sm:sec:harm} derives the angular-harmonic model of
Eqs.~(14)--(19) and states its assumptions. Notation follows the main text.
$W_{n\bk,n'\bk'}$ is the rate from $|n\bk\rangle$ to $|n'\bk'\rangle$,
$L=L^{\mathrm{ph}}+L^{\mathrm{sym}}+L^{\mathrm{skew}}$, and $\Nimp$ is the
defect fraction per sulfur site. Equations of the main text are cited by their
number there. Equations of this document carry the prefix S.

\section{Steady photocurrent without recombination}
\label{sm:sec:steady}

The recombination-free Boltzmann equation conserves carrier number under
collisions, while optical generation continually adds carriers. The total
population therefore grows with time. The current can still approach a
steady value if this growth is confined to collision-invariant modes that
carry zero current. We give sufficient conditions below.

\subsection{Setting}

Consider one dilute carrier sector on a finite, time-reversal-paired grid,
with state indices including bands where needed. The master equation is
\begin{equation}
 \partial_t\delta f_{\bk}=G_{\bk}-(L\delta f)_{\bk},\qquad
 (L g)_{\bk}=\nu_{\bk}g_{\bk}-\sum_{\bk'}W_{\bk'\bk}g_{\bk'},\qquad
 \nu_{\bk}=\sum_{\bk'\ne\bk}W_{\bk\bk'}.
 \label{sm:eq:A-bte}
\end{equation}
Here $g$ is an arbitrary vector on the discrete carrier-state grid, with
components $g_{\bk}$, and $\nu_{\bk}$ is the total out-scattering rate from
state $\bk$. The source $G_{\bk}$ is the optical generation rate, and
$\delta f_{\bk}$ is the photoinduced occupation.
Self-transitions are omitted, and the total transition rates must satisfy
$W_{\bk\bk'}\ge0$. Although the skew contribution can have either sign,
the truncated Born expansion must remain in a regime where the total rates
are nonnegative. This dilute-carrier equation neglects Pauli blocking and assumes
an equilibrium phonon bath~\cite{SMJacoboni2010,SMBernardi2016,SMZhou2021}.

The following three properties of the collision operator give a simple sufficient criterion:

\noindent
(i) $\mathbf1^{\mathsf T}L=0$ (particle conservation), as follows by summing
Eq.~\eqref{sm:eq:A-bte}, where $\mathbf1$ is the vector with every component
equal to one.

\noindent
(ii) The directed transition graph has a single closed communicating class
$\mathcal C$. This is a set of states that can reach one another through
scattering, with every allowed transition from the set staying inside it.
All other states are transient: under collisions alone, carriers starting
outside $\mathcal C$ eventually enter it with probability one.

\noindent
(iii) Its stationary zero mode $\phi_0$ is even under time reversal,
$\phi_{0,-\bk}=\phi_{0,\bk}$.

\subsection{Spectrum of the collision operator}
\label{sm:sec:spec}

The nonnegative transition rates and particle conservation make $-L$ a
finite continuous-time Markov generator. Under property (ii), we apply
the Perron--Frobenius theorem within the closed class $\mathcal C$.
It gives a unique zero mode within that class. The population on the
transient states decays into $\mathcal C$ under collisions alone.
Together, these results give one zero eigenvalue of $L$, with algebraic
multiplicity one. The normalized zero mode $\phi_0$ is positive within
$\mathcal C$ and zero outside it. Every other eigenvalue has positive real
part, so all remaining modes decay~\cite{SMNorris1997}.

The spectral argument uses particle conservation and one closed class rather
than pairwise detailed balance~\cite{SMNorris1997}. At finite temperature, the Boltzmann distribution
$\phi_{0,\bk}\propto e^{-\epsilon_{\bk}/k_BT}$ is stationary under
$L^{\mathrm{ph}}$ and $L^{\mathrm{sym}}$ by detailed balance~\cite{SMGiustino2017,SMZhou2021}, and under
$\Lsk$ by Eq.~\eqref{sm:eq:Lf0}. It depends on $\bk$ only through
$\epsilon_{\bk}$ and is therefore even. Here $\epsilon_{\bk}$ is the carrier
energy, $T$ is the temperature, and $k_B$ is Boltzmann's constant.

At strictly $0$~K, the equilibrium Bose occupation $n_B$ of the phonon modes
vanishes, removing phonon absorption. Emission lowers carrier
energy, while elastic defects preserve it. Scattering can therefore connect
high-energy states to lower-energy states through one-way paths.
Suppose every retained state can reach the same terminal band-edge shell.
Also suppose elastic scattering connects all states within that shell,
including any distinct valleys. Here ``terminal'' means that every allowed
scattering transition from the shell stays within it. The shell is then the
unique closed class $\mathcal C$, and the other states are transient.
This establishes (ii), with $\phi_0$ supported on the shell.

For a complete, time-reversal-paired shell, a uniform population is stationary
under elastic symmetric scattering. It is also stationary under skew
scattering by Eq.~\eqref{sm:eq:Lf0}. With absorption absent and no allowed
lower-energy final state, phonons leave this population unchanged.
The population is even under time reversal,
which establishes (iii) in this setting.

\subsection{Decomposition of the distribution}
\label{sm:sec:decomp}

Under property (ii), normalize $\mathbf1^{\mathsf T}\phi_0=1$ and define
\begin{equation}
 P=\phi_0\mathbf1^{\mathsf T},\qquad Q=I-P,\qquad
 H=\operatorname{ran}Q=\{g:\mathbf1^{\mathsf T}g=0\}.
 \label{sm:eq:A-projectors}
\end{equation}
Here $I$ is the identity operator on the carrier-state space. The projector
$P$ extracts the component along the zero mode, while
$Q$ extracts a deviation with zero net carrier number. For a nonuniform
$\phi_0$, $P$ is an oblique projector because $\phi_0$ is not parallel to
$\mathbf1$. Particle conservation and $L\phi_0=0$ give
\[
 PL=LP=0,\qquad QL=LQ=L.
\]
In particular, $L$ maps $H$ into itself. The notation $L_H=L|_H$ denotes
the restriction of $L$ to this invariant subspace. Its action on any
$g\in H$ is simply $L_Hg=Lg$. Since the only zero mode is $\phi_0$ and
$\mathbf1^{\mathsf T}\phi_0=1$, no nonzero vector in $H$ lies in the
kernel of $L$. Thus $L_H$ is invertible on the finite-dimensional space $H$.

We now solve Eq.~\eqref{sm:eq:A-bte} for a time-independent optical source
$G$. Using the notation of Eq.~(12) in the main text, write
\[
 \delta f(t)=a(t)\phi_0+\widetilde f(t),\qquad
 a(t)=\mathbf1^{\mathsf T}\delta f(t),\qquad \widetilde f(t)=Q\delta f(t)\in H.
\]
Summing the kinetic equation over all states gives
\[
 \dot a(t)=\mathbf1^{\mathsf T}G\equiv R_{\mathrm{gen}},\qquad
 a(t)=a(0)+tR_{\mathrm{gen}}.
\]
The zero-mode contribution is therefore
$a(t)\phi_0=P\delta f(0)+tR_{\mathrm{gen}}\phi_0$.
Applying $Q$ to the same kinetic equation gives the remaining evolution,
\[
 \frac{d\widetilde f(t)}{dt}=QG-L_H\widetilde f(t),\qquad
 \widetilde f(0)=Q\delta f(0).
\]
Its stationary solution is $\widetilde f^{\mathrm{ss}}=L_H^{-1}QG$. Subtracting this
solution yields a homogeneous equation,
\[
 \frac{d}{dt}\big[\widetilde f(t)-\widetilde f^{\mathrm{ss}}\big]
 =-L_H\big[\widetilde f(t)-\widetilde f^{\mathrm{ss}}\big].
\]
Hence
\[
 \widetilde f(t)=L_H^{-1}QG+
 e^{-L_Ht}\big[Q\delta f(0)-L_H^{-1}QG\big].
\]
Combining these expressions for $a(t)$ and $\widetilde f(t)$ gives the
explicit solution of the decomposition in Eq.~(12) of the main text,
\begin{equation}
 \delta f(t)=P\delta f(0)+tR_{\mathrm{gen}}\phi_0+
 L_H^{-1}QG+e^{-L_Ht}\big[Q\delta f(0)-L_H^{-1}QG\big].
 \label{sm:eq:A-main}
\end{equation}
Here the last two terms are embedded in $H$, and the last term decays as
established in
Sec.~\ref{sm:sec:spec}.

\subsection{Finiteness of the current}

The number-current functional is $j[g]=\sum_{\bk}\bv_{\bk}g_{\bk}$.
Time reversal gives $\bv_{-\bk}=-\bv_{\bk}$, so property (iii) implies
\begin{equation}
 j[\phi_0]=\frac12\sum_{\bk}
 \big(\bv_{\bk}\phi_{0,\bk}+\bv_{-\bk}\phi_{0,-\bk}\big)=0.
 \label{sm:eq:A-parity}
\end{equation}
Thus the growing term in Eq.~\eqref{sm:eq:A-main} contributes no current and
\begin{equation}
 j(t)\longrightarrow j^{\mathrm{ss}}=j[L_H^{-1}QG].
 \label{sm:eq:A-jss}
\end{equation}
This establishes the finite-current statement used after Eq.~(13) of the
main text, including at $0$~K when the stated conditions hold. The proof
uses particle conservation, one closed class, and an even zero mode.
Although skew scattering can convert a generic even distribution into an
odd component, collisions leave $\phi_0$ unchanged because it is a zero mode
of the full operator $L$.

\subsection{Electron and hole sectors}
\label{sm:sec:eh}

Without recombination or interband scattering, the electron and hole
collision operators form separate blocks. Generation adds equal numbers of
electrons and holes, and the preceding argument applies to each dilute
carrier block when it satisfies properties (i)--(iii). For each sector
$s=c,h$, denoting conduction electrons and holes, respectively, let
$g^s=\widetilde f^{s,\mathrm{ss}}=(L_H^s)^{-1}Q^sG^s$ be its stationary
non-growing distribution. The superscript $s$ labels the corresponding
sector's operator, projector, and source. The physical charge current is
\begin{equation}
 J^{\mathrm{ss}}=-e\sum_{\bk}\bv^c_{\bk}g^c_{\bk}
                  +e\sum_{\bk}\bv^h_{\bk}g^h_{\bk},\qquad e>0.
 \label{sm:eq:A-eh}
\end{equation}
Here $e$ is the positive elementary charge, and $\bv^c_{\bk}$ and
$\bv^h_{\bk}$ are the electron and hole band velocities. The hole momentum
is opposite to that of the missing valence electron.
Equivalently, using valence-electron deviations
$\delta f^v_{\bk}=-g^h_{-\bk}$ and
$\bv^h_{-\bk}=\bv^v_{\bk}$ gives the usual electron-language charge current.
The electron and hole contributions generally do not cancel, so the main
text reports their signed sum as the total response.

\subsection{Regime of validity}
\label{sm:sec:validity}

Real carriers cannot accumulate indefinitely. Both the stationary
anisotropic deviation and the growing population must remain dilute and
unsaturated~\cite{SMJacoboni2010,SMBernardi2016}.
Let $\tau_J$ be the decay time of the current-carrying modes and
$\tau_{\mathrm{rec}}$ the recombination time. Let
$\mathcal N_{\mathrm{avail}}$ denote the population available before band
filling or depletion invalidates the linear source. A measurable plateau
requires the time window
\begin{equation}
 \tau_J\ll t\ll
 \min\{\tau_{\mathrm{rec}},\mathcal N_{\mathrm{avail}}/R_{\mathrm{gen}}\}.
 \label{sm:eq:A-window}
\end{equation}
A large elastic out-scattering rate does not by itself guarantee small
occupations, because energy relaxation can be much slower and allow carriers
to accumulate in long-lived states. Within the linear regime,
$G\propto I_{\mathrm{light}}$ implies $J^{\mathrm{ss}}\propto I_{\mathrm{light}}$,
where $I_{\mathrm{light}}$ is the illumination intensity.
\clearpage
\section{Angular-harmonic model}
\label{sm:sec:harm}

\subsection{Collision operator, angular harmonics, and current}

Under continuous-wave illumination in the low-injection limit, the photoinduced
electron or hole occupation $\delta f_{n\bk}=f_{n\bk}-f^{0}_{n\bk}$, measured
relative to the equilibrium occupation $f^{0}_{n\bk}$, obeys the linear
dilute-carrier Boltzmann equation of the main text,
\begin{equation}
  \partial_t\,\delta f = G - L\,\delta f,
  \qquad
  L = L^{\mathrm{ph}} + L^{\mathrm{sym}} + \Lsk .
  \label{sm:eq:bte}
\end{equation}
Each collision operator has the out-minus-in form
\begin{equation}
  (L^{x}\delta f)_{\bk}
  = \sum_{\bk'}\big[W^{x}_{\bk\bk'}\,\delta f_{\bk}
                  - W^{x}_{\bk'\bk}\,\delta f_{\bk'}\big],
  \qquad x\in\{\mathrm{ph},\mathrm{sym},\mathrm{skew}\},
  \label{sm:eq:outin}
\end{equation}
where $W^{x}_{\bk\bk'}$ is the rate from $\bk$ to $\bk'$. Band indices are
suppressed, and the model treats one carrier channel at a time. We separate
ordinary relaxation from elastic skew scattering,
\begin{equation}
  \Lsym \equiv L^{\mathrm{ph}}+L^{\mathrm{sym}},
  \qquad
  \Wsk_{\bk\bk'} = -\Wsk_{\bk'\bk}.
  \label{sm:eq:split}
\end{equation}
The superscript $(S)$ is reserved for the combined ordinary operator
$L^{\mathrm{ph}}+L^{\mathrm{sym}}$. The elastic symmetric defect kernel and
operator are denoted by $W^{\mathrm{sym}}$ and $L^{\mathrm{sym}}$ throughout.
The antisymmetry
of $\Wsk$ follows from its definition
$\Wsk=\tfrac12(W-W^{\mathsf T})$ in the main text.

Defect scattering is elastic, so $L^{\mathrm{sym}}$ and $\Lsk$ connect only
states on the same constant-energy contour. Phonon scattering connects different
energy shells. The model keeps it only as a loss rate from a given shell, as
described in Sec.~\ref{sm:sec:sym}. The steady state can then be solved shell by
shell, and the current is obtained by summing over shells.

\medskip
\noindent\textbf{Angular harmonics.}

For the reduced model, take an effective circular energy contour with a
uniform angular measure, parameterized by $\theta$, avoiding the nonuniform
measure and additional pocket indices that general contours may require.
We expand
\begin{equation}
  \delta f(\theta) = \sum_{m\in\mathbb Z} f_m\, e^{im\theta},
  \qquad f_{-m}=f_m^{*}.
  \label{sm:eq:harm}
\end{equation}
This angular-harmonic expansion is the two-dimensional analogue of a
Legendre expansion and uses a complete basis $e^{im\theta}$ on the chosen
circle, which approximates the actual constant-energy contour. Since
$\delta f$ is real, its coefficients satisfy $f_{-m}=f_m^*$.

The angle is measured from the $k_x$ axis about the valley center. For the
conduction band the center is $\Gamma$. For the valence band the shell consists
of the two time-reversal-related valleys at $\pm\bk_0$ on $\Gamma$--Y. Since $\bk\to-\bk$ exchanges these valleys, the parity statements below
refer to their combined distribution on an effective time-reversal-paired
shell. The reduced model represents this operation by $\theta\to\theta+\pi$.

The index $m$ counts the lobes of the angular pattern. $m=0$ is isotropic,
$m=\pm1$ is a one-sided displacement, and $m=\pm2$ is a two-lobed alignment.
Under $\theta\to\theta+\pi$, which represents $\bk\to-\bk$ in the effective
paired-shell description, even-$m$ patterns remain unchanged, whereas
odd-$m$ patterns change sign.

\medskip
\noindent\textbf{Current.}

The band velocity is odd in $\bk$ and, in the isotropic approximation used
below, takes the form $\bv(\theta)=v\,(\cos\theta,\sin\theta)$, containing only
$m=\pm1$ harmonics. Here $v$ is the speed on the chosen circular energy
shell. Projecting the distribution onto this velocity and using
harmonic orthogonality gives
\begin{equation}
  J_x \propto \int\!\frac{d\theta}{2\pi}\,\cos\theta\,\delta f(\theta)
      = \Re f_1,
  \qquad
  J_y \propto \int\!\frac{d\theta}{2\pi}\,\sin\theta\,\delta f(\theta)
      = -\Im f_1 .
  \label{sm:eq:Jf1}
\end{equation}
Within this isotropic velocity approximation only $f_1$ carries current,
whereas anisotropic velocities can also project onto higher odd harmonics. An $m=2$ alignment is anisotropic, but it has equal
weight at $\theta$ and $\theta+\pi$ and so carries no current.

\subsection{Symmetry selection rules}
\label{sm:sec:selection}

For elastic scattering by nonmagnetic static defects, the following symmetry
relations follow from time reversal and exchange~\cite{SMIsobeXuFu2020,SMDeyo2009}.
Expand the defect kernel on the effective shell as
$W(\theta,\theta')=\sum_{\mu\nu}c_{\mu\nu}\,e^{i\mu\theta}e^{i\nu\theta'}$,
where $\mu,\nu\in\mathbb Z$ are harmonic indices and $c_{\mu\nu}$ are the
corresponding Fourier coefficients.

The first symmetry is microscopic reversibility. In a nonmagnetic crystal, time
reversal gives $W_{\bk\bk'}=W_{-\bk',-\bk}$, which reads
$W(\theta,\theta')=W(\theta'+\pi,\theta+\pi)$ on the shell. In terms of the
coefficients,
\begin{equation}
  c_{\mu\nu} = (-1)^{\mu+\nu}\,c_{\nu\mu}.
  \label{sm:eq:TR}
\end{equation}

The second symmetry is the exchange property. The elastic symmetric defect
kernel $W^{\mathrm{sym}}$ has coefficients
$c^{\mathrm{sym}}_{\mu\nu}=c^{\mathrm{sym}}_{\nu\mu}$, and the skew part obeys
$c^{\mathrm{skew}}_{\mu\nu}=-c^{\mathrm{skew}}_{\nu\mu}$. Combining with
Eq.~\eqref{sm:eq:TR},
\begin{equation}
  c^{\mathrm{sym}}_{\mu\nu}\neq0 \ \Rightarrow\ \mu+\nu\ \text{even},
  \qquad
  c^{\mathrm{skew}}_{\mu\nu}\neq0 \ \Rightarrow\ \mu+\nu\ \text{odd}.
  \label{sm:eq:parity}
\end{equation}
The in-scattering term acting on $e^{im\theta'}$ samples coefficients
$c_{\mu,-m}$ (or their exchanged counterparts). The out-scattering term
satisfies the same parity restriction. Equation~\eqref{sm:eq:parity} is
therefore a parity rule for the elastic defect kernel. Symmetric defect scattering connects harmonics of equal
parity, with $\mu-m$ even. Skew scattering connects harmonics of opposite parity,
with $\mu-m$ odd. The symmetric operator relaxes the even and odd sectors
separately and can never create current from an even distribution. The leading phonon relaxation channel in an equilibrium time-reversal-invariant
bath also preserves global $\bk$ parity. Within the population-current
mechanisms retained here, skew scattering couples the even source to the odd
current sector. Thus $\Lsym$ preserves global momentum parity.

Two consequences are used below. First, a rotationally invariant kernel,
$W=W(\theta-\theta')$, has $c_{\mu\nu}=w_\mu\delta_{\nu,-\mu}$, where
$w_\mu=\int_0^{2\pi}d\phi\,W(\phi)e^{-i\mu\phi}/(2\pi)$ is its Fourier
coefficient and $\delta_{\nu,-\mu}$ is the Kronecker delta. Thus $\mu+\nu=0$
is always even. By Eq.~\eqref{sm:eq:parity}, a rotationally invariant skew kernel
is forbidden. The skew kernel must carry the orientation of the crystal, here
the polar axis of $mm2$. This is what allows it to convert an alignment into a
directed current.

Second, the mirror $m_x$ of $mm2$ acts as $\theta\to\pi-\theta$ and maps
$f_m\to(-1)^m f_{-m}$. A mirror-symmetric distribution has $f_{-1}=-f_1$, so
$f_1$ is purely imaginary. By Eq.~\eqref{sm:eq:Jf1}, $J_x=0$ while $J_y\neq0$.
This is the angular-basis form of the statement in the main text that
$\sigma^{yyy}$ is the only in-plane component under $y$-polarized light.

\subsection{Angular-harmonic lifetimes}
\label{sm:sec:sym}

We first consider the elastic symmetric defect kernel. In the isotropic
approximation it depends only on the scattering angle
$\phi=\theta-\theta'$, so
$W^{\mathrm{sym}}(\theta,\theta')=W^{\mathrm{sym}}(\phi)$ and
$W^{\mathrm{sym}}(\phi)=W^{\mathrm{sym}}(-\phi)$.
This is the angular-relaxation construction used in transport
theory~\cite{SMAshcroftMermin1976,SMKoniakhin2014}. The skew kernel remains
anisotropic, as required by Sec.~\ref{sm:sec:selection}.
The symmetric defect part of Eq.~\eqref{sm:eq:outin} becomes a convolution.
Its eigenfunctions are the angular harmonics,
\begin{equation}
  L^{\mathrm{sym}}e^{im\theta}=\alpha_m\Nimp e^{im\theta},
  \qquad
  \alpha_m\Nimp=\int\!\frac{d\phi}{2\pi}\,W^{\mathrm{sym}}(\phi)\,\big(1-\cos m\phi\big).
  \label{sm:eq:taum}
\end{equation}
The $\sin m\phi$ part integrates to zero because $W^{\mathrm{sym}}$ is even. Each
harmonic relaxes with its own rate, weighted by $1-\cos m\phi$. For $m=1$ this
is the usual momentum-relaxation rate, which is insensitive to small-angle
scattering. For $m=2$ it is the alignment-relaxation rate, which is most
sensitive to scattering near $90^\circ$. For narrowly forward scattering,
the approximation $1-\cos m\phi\simeq m^2\phi^2/2$ makes the defect relaxation
rate proportional to $m^2$, so the $m=1$ and $m=2$ defect-limited lifetimes
differ by a finite factor rather than being parametrically separated in
this limit.

Since independent phonon and defect transition kernels add, we represent
phonon scattering in the reduced model by a diagonal effective decay rate
$\Gamma_m(T)$, including loss to other shells that is absent from the elastic
convolution in Eq.~\eqref{sm:eq:taum}. The total relaxation rate is then
\begin{equation}
  \tau_m^{-1} = \Gamma_m(T) + \alpha_m\,\Nimp .
  \label{sm:eq:rate-split}
\end{equation}
Here $W^{\mathrm{sym}}=\Nimp W^{\mathrm{sym}}_1$, where
$W^{\mathrm{sym}}_1$ is the single-defect kernel of Eq.~(4) of the main text.
Equation~\eqref{sm:eq:taum} therefore gives the defect rate $\alpha_m\Nimp$.
Linearity in $\Nimp$ holds for dilute,
uncorrelated defects, without multiple-scattering interference~\cite{SMLu2019,SMLu2020}. This is also
the regime in which the single-defect rates were rescaled in the numerics.
$\Gamma_m(T)$ is the phonon-limited rate of harmonic $m$. It contains the
inelastic loss from the shell, which is common to all $m$. It also includes
angular relaxation from quasi-elastic phonon scattering within the shell.
The model places the temperature dependence in $\Gamma_m(T)$ and assumes
that cooling reduces these rates for a fixed carrier distribution. The defect rates $\alpha_m$ and the skew coefficients introduced
below are taken independent of $T$ and of $\Nimp$. This means that the defect
potential and its screening are those of the fixed, undoped host.

For $m=0$ the weight $1-\cos m\phi$ vanishes. Elastic scattering therefore does
not relax the isotropic occupation of a shell, and $\alpha_0=0$ exactly. The
value of $f_0$ on a given shell is set by energy relaxation and generation. This mode
plays no role in the current, as shown in Sec.~\ref{sm:sec:skew}, so its value is
not needed.

\subsection{Optical source}
\label{sm:sec:source}

For single-photon electric-dipole absorption, the lowest-order population
source is~\cite{SMSipeShkrebtii2000}
\begin{equation}
  G_{c\bk} \propto \big|\be\cdot\bp_{cv}(\bk)\big|^2\,
  \delta\!\big(\epsilon_{c\bk}-\epsilon_{v\bk}-\hbar\omega\big),
  \label{sm:eq:G}
\end{equation}
with $\bp_{cv}(\bk)=\langle c\bk|\hat\bp|v\bk\rangle$. For time-reversed partner states, time reversal gives
$\epsilon_{n,-\bk}=\epsilon_{n\bk}$ and, up to Bloch-state phases,
$\bp_{cv}(-\bk)=-\bp_{cv}(\bk)^{*}$.
The sign and the conjugation disappear in the modulus squared. For a real
polarization vector $\be$, which is linearly polarized light, this gives
$G(-\bk)=G(\bk)$. The source is globally even (with partner bands summed when needed). In the
effective paired-shell representation its expansion contains only even
harmonics,
\begin{equation}
  G(\theta)=\sum_{m\ \mathrm{even}} G_m e^{im\theta},
  \qquad G_1=0 .
  \label{sm:eq:G1zero}
\end{equation}
The same conclusion holds in the length gauge, where
$\bp_{cv}\to i m_e\omega_{cv}\,\mathbf r_{cv}$ and
$\mathbf r_{cv}(-\bk)=+\mathbf r_{cv}(\bk)^*$. Here $m_e$ is the electron
mass, $\omega_{cv}=(\epsilon_{c\bk}-\epsilon_{v\bk})/\hbar$ is the interband
transition frequency, and $\mathbf r_{cv}$ is the interband position matrix
element. No odd population is injected directly at this order. This excludes the
corresponding direct injection current for linearly polarized light, but not
the intrinsic shift current, which is a distinct optical response~\cite{SMSipeShkrebtii2000,SMYoungRappe2012}.
Circular polarization or phonon-assisted ballistic generation can add odd
source terms~\cite{SMDai2021} and are not considered here.

As an illustrative circular-contour source, assume that the transition dipole
rotates with $\bk$,
$\bp_{cv}(\theta)\propto(\cos\theta,\sin\theta)$. For $\be=\hat y$,
\begin{equation}
  |\be\cdot\bp_{cv}|^2\propto\sin^2\theta=\tfrac12-\tfrac12\cos2\theta .
  \label{sm:eq:sin2}
\end{equation}
With the complex angular-harmonic convention of Eq.~\eqref{sm:eq:G1zero}, this source
has $G_{+2}=G_{-2}=-G_0/2$, or a real cosine amplitude of $-G_0$, with this
normalization set by the illustrative dipole above. The $\bk$ dependence of the true matrix elements and the
anisotropy of the contours add higher even harmonics $G_4,G_6,\dots$ to
Eq.~\eqref{sm:eq:sin2}. These are neglected in the model and discussed in
Sec.~\ref{sm:sec:beyond}.

\subsection{Skew conversion}
\label{sm:sec:skew}

Inserting the antisymmetry $\Wsk_{\bk'\bk}=-\Wsk_{\bk\bk'}$ into
Eq.~\eqref{sm:eq:outin} merges the two terms,
\begin{equation}
  (\Lsk\delta f)_{\bk} = \sum_{\bk'}\Wsk_{\bk\bk'}\,
  \big(\delta f_{\bk}+\delta f_{\bk'}\big).
  \label{sm:eq:skewsum}
\end{equation}
Elastic symmetric scattering acts on occupation differences. The skew term
acts on their sum and converts angular components. When its row sums vanish
as shown below, its real matrix is antisymmetric and contributes no quadratic
dissipation on the elastic shell, while still changing individual occupations~\cite{SMIsobeXuFu2020}.

\medskip
\noindent\textbf{The isotropic population generates no current.}

For a constant $\delta f=f_0$, Eq.~\eqref{sm:eq:skewsum} reduces to
$(\Lsk f_0)_{\bk}=2f_0\,\Sigma_{\bk}$ with
$\Sigma_{\bk}\equiv\sum_{\bk'}\Wsk_{\bk\bk'}$. Antisymmetry alone gives only
$\sum_{\bk}\Sigma_{\bk}=0$, which is particle conservation. It does not give
$\Sigma_{\bk}=0$ for each $\bk$. The parity rule \eqref{sm:eq:parity} likewise
permits $m=0\to m=1$. The vanishing follows instead from the explicit form of
the skew rate, Eq.~(6) of the main text,
\begin{equation}
  \Wsk_{\bk\bk'} \propto \sum_{\bk''}
  T(\bk,\bk',\bk'')\,
  \delta(\epsilon_{\bk}-\epsilon_{\bk'})\,\delta(\epsilon_{\bk}-\epsilon_{\bk''}),
  \qquad
  T(\bk,\bk',\bk'')\equiv\Im\!\big[U_{\bk\bk'}U_{\bk'\bk''}U_{\bk''\bk}\big].
  \label{sm:eq:Wsk-explicit}
\end{equation}
Because $U=U^{\dagger}$, exchanging any two arguments of the triple product
turns it into its complex conjugate,
\begin{equation}
  U_{\bk\bk''}U_{\bk''\bk'}U_{\bk'\bk}
  =\big(U_{\bk\bk'}U_{\bk'\bk''}U_{\bk''\bk}\big)^{*},
  \qquad\text{so}\qquad
  T(\bk,\bk'',\bk')=-T(\bk,\bk',\bk'') .
  \label{sm:eq:T-antisym}
\end{equation}
Likewise $T(\bk',\bk,\bk'')=-T(\bk,\bk',\bk'')$, which is the antisymmetry
$\Wsk_{\bk\bk'}=-\Wsk_{\bk'\bk}$ used throughout. Summing
Eq.~\eqref{sm:eq:Wsk-explicit} over $\bk'$ gives
\begin{equation}
  \Sigma_{\bk}\propto\sum_{\bk',\bk''}
  T(\bk,\bk',\bk'')\,
  \delta(\epsilon_{\bk}-\epsilon_{\bk'})\,\delta(\epsilon_{\bk}-\epsilon_{\bk''}) .
  \label{sm:eq:hermit}
\end{equation}
The two intermediate states now enter on an equal footing. The summation domain
and the product of delta functions are symmetric under
$\bk'\leftrightarrow\bk''$, while $T$ is antisymmetric by
Eq.~\eqref{sm:eq:T-antisym}. The sum therefore vanishes in pairs, and
$\Sigma_{\bk}=0$ for every $\bk$. Hence
\begin{equation}
  \Lsk f_0 = 0
  \label{sm:eq:Lf0}
\end{equation}
for any energy-only $f_0$, at the retained third order for a static Hermitian
defect potential and complete, consistently weighted intermediate-state sums,
without assuming isotropy. This is the elastic probability-sum
constraint underlying the skew collision integral~\cite{SMIsobeXuFu2020}.
Thus skew scattering leaves the isotropic population unchanged but can act on the anisotropic part of the distribution.

\medskip
\noindent\textbf{Conversion of the alignment.}

The parity rule allows the retained alignment to couple to $m=1$, but the
coupling strength is material dependent and may vanish. We define it through
the projection of the skew collision operator below.
In the mirror-symmetric alignment sector, $f_{-2}=f_2$ and $G_{-2}=G_2$.
Together with the reality of the distribution and source, these relations
make $f_2$ and $G_2$ real, so we use them directly as in the main text.
For the current projection, define $a_1=-\Im f_1$, so $J_y\propto a_1$.
For the mirror-symmetric alignment sector, define the real coefficient $S_2$
by
\begin{equation}
 -\Im\big(\Lsk\delta f^{(2)}\big)_1
       =-S_2\Nimp f_2,\qquad
 \delta f^{(2)}(\theta)=f_2(e^{2i\theta}+e^{-2i\theta}).
 \label{sm:eq:S2}
\end{equation}
The subscript $1$ selects the $m=1$ Fourier coefficient, and $S_2$ includes
both $m=\pm2$ components. Its sign depends on the defect potential,
the source orientation, and the current convention. In the dilute
single-defect Born expansion, the skew rate is linear in defect fraction and
third order in the potential, whereas the leading symmetric rate is second
order~\cite{SMIsobeXuFu2020,SMLu2019}.

The first-order skew approximation requires the norm of
$(L^{(S)})^{-1}\Lsk$ on the relaxing subspace to be small. Under this condition,
the odd response is first order in skew strength and its feedback on the even
alignment is second order. Higher odd harmonics do not contribute to the
isotropic-velocity current at this order. The alignment $f_2$ and source $G_2$ are the same amplitudes used in
Eqs.~(17)--(18) of the main text, with fixed current-normalization factors
absorbed into the proportionality constant.

\subsection{Closed-form response and relation to the full calculation}
\label{sm:sec:closed}
\label{sm:sec:beyond}

Balancing each harmonic in Eq.~\eqref{sm:eq:bte} at steady state, and using
Eqs.~\eqref{sm:eq:rate-split}, \eqref{sm:eq:G1zero}, \eqref{sm:eq:Lf0} and
\eqref{sm:eq:S2}, gives
\begin{align}
  f_2 &= \tau_2\,G_2 = \frac{G_2}{\Gamma_2(T)+\alpha_2\Nimp},
  \label{sm:eq:f2}\\[2pt]
  a_1 &= \tau_1\,S_2\Nimp\,f_2
       = \frac{S_2\Nimp}{\Gamma_1(T)+\alpha_1\Nimp}\;
         \frac{G_2}{\Gamma_2(T)+\alpha_2\Nimp}.
  \label{sm:eq:f1}
\end{align}
With Eq.~\eqref{sm:eq:Jf1}, and after dividing by the fixed illumination
intensity,
\begin{equation}
  J(\Nimp,T)\ \propto\
  \frac{S_2\,G_2\,\Nimp}
       {\big[\Gamma_1(T)+\alpha_1\Nimp\big]\big[\Gamma_2(T)+\alpha_2\Nimp\big]} .
  \label{sm:eq:J}
\end{equation}
This is Eq.~(18) of the main text. The defect fraction appears once in the
numerator, through the conversion, and twice in the denominator, through the
relaxation of the alignment that is converted and of the current that results.

With $S_2$, $G_2$, and $\alpha_m$ held fixed, the finite optimal fraction in
Eq.~(19) of the main text requires positive $\Gamma_m$ and $\alpha_m$ and a
nonzero $S_2G_2$. At strictly $0$~K, spontaneous emission can vanish at a band
minimum~\cite{SMGiustino2017}, in which case the low-fraction limit must be
re-examined under the steady-current conditions of Sec.~\ref{sm:sec:steady}.

\medskip
\noindent\textbf{Relation to the full calculation.}

The model uses a circular effective shell, isotropic ordinary relaxation, a
truncated source, diagonal effective phonon losses, and weak-skew conversion.
The full calculation retains the actual bands, angular structure, and
in-scattering rather than these shell-model reductions. The isotropic
approximation makes $L^{\mathrm{sym}}$ diagonal in the angular basis,
Eq.~\eqref{sm:eq:taum}. Replacing the phonon operator by diagonal effective
losses then gives one $\tau_m$ per harmonic, Eq.~\eqref{sm:eq:rate-split}.
Another approximation is the truncation of the even sector to
$m=0,2$, Eq.~\eqref{sm:eq:sin2}. The underlying global momentum-parity rules, the energy-only cancellation
\eqref{sm:eq:Lf0}, source evenness, and the mirror constraint on current
direction survive under their stated physical assumptions. The
full solution solves Eq.~\eqref{sm:eq:bte} with the anisotropic $\Lsym$, the full
skew operator, and all bands, and retains the complete in-scattering structure.
The model is used to interpret the single-channel scaling, namely the
unit-slope rise, the plateau or peak, the optimal fraction in Eq.~(19) of the
main text, and the cooling enhancement in Fig.~2. Absolute magnitudes and the signed sum of
the electron and hole channels are taken from the numerics.

Structure beyond the model can arise from the neglected even harmonics. The
anisotropic matrix elements of GeS generate $G_4,G_6,\dots$, and anisotropic
symmetric scattering mixes them with $G_2$. Each even harmonic $m$ is converted
to $m=1$ by $\Lsk$ with its own coefficient $S_m$, defined in the same
mirror-symmetric sector and phase convention as $S_2$ by
\[
 -\Im\big(\Lsk\delta f^{(m)}\big)_1=-S_m\Nimp f_m,\qquad
 \delta f^{(m)}(\theta)=f_m(e^{im\theta}+e^{-im\theta}),\qquad m=2,4,\ldots .
\]
Here $f_m$ is real in this sector. Each alignment is quenched at its own threshold
$\Gamma_m/\alpha_m$. Two such pathways with conversion coefficients
of opposite sign and different thresholds can produce a sign reversal if their
relative weights exchange dominance as $\Nimp$ varies. This is one mechanism consistent with
the behavior of the Se$_{\mathrm S}$ hole channel at $300$~K in the main text.

\par
\endgroup
\pagestyle{supplement}

  \begingroup
  \renewcommand{\bibsection}{\section*{Supplemental References}}
  
  \endgroup

\SMEndDocument

\end{document}